\documentclass[preprint,english,showpacs,12pt,floatfix]{revtex4-1}
\usepackage{babel,amsmath,amssymb,dcolumn}
\usepackage{hyperref}
\usepackage{amsfonts}
\usepackage{amssymb}
\usepackage{graphicx}
\usepackage{graphics}
\usepackage{latexsym}
\usepackage{dcolumn}
\usepackage{epsfig}
\usepackage{subfigure}
\usepackage{array}
\usepackage{multirow}
\usepackage[retainorgcmds]{IEEEtrantools}
\allowdisplaybreaks[1]
\begin{document}

\newcommand{\vp}{\varphi}
\newcommand{\nn}{\nonumber\\}
\newcommand{\beq}{\begin{equation}}
\newcommand{\eeq}{\end{equation}}
\newcommand{\bed}{\begin{displaymath}}
\newcommand{\eed}{\end{displaymath}}
\def\bea{\begin{eqnarray}}
\def\eea{\end{eqnarray}}
\newcommand{\veps}{\varepsilon}
\newcommand{\nablasl}{{\slash \negthinspace \negthinspace \negthinspace \negthinspace  \nabla}}
\newcommand{\om}{\omega}

\newcommand{\Dsl}{{\slash \negthinspace \negthinspace \negthinspace \negthinspace  D}}
\newcommand{\tDsl}{{\tilde \Dsl}}
\newcommand{\tnablasl}{{\tilde \nablasl}}
\title{Vacuum polarization of the quantized massive scalar field in the global monopole spacetime I: the field fluctuation}

\author{Owen Pavel Fern\'{a}ndez Piedra$^{1, 2}$ \\
\textit{$^1$ Departamento de F\'isica, Divisi\'on de Ciencias e Ingenier\'ias, Universidad de Guanajuato, Campus Le\'on, Loma del Bosque N0. 103, Col. Lomas del Campestre, CP 37150, Le\'on, Guanajuato, M\'exico.}\\
\textit{$^2$Grupo de Estudios Avanzados, Universidad de Cienfuegos, Carretera a Rodas, Cuatro Caminos, s/n. Cienfuegos, Cuba.}}
\email{opavelfp2006@gmail.com}

\begin{abstract}
We study the vacuum polarization of a massive scalar field $\phi$ with arbitrary coupling to gravity in pointlike global monopole spacetime. Using Schwinger-DeWitt proper time formalism, we calculate the vacuum expectation value $<\phi^{2}>$, when the Compton length of the quantum field is much less than the characteristic radius of the curvature of the background geometry, and we can ignore nonlocal contributions. Explicit analytic expressions are obtained for a general curvature coupling parameter, and specified to the more physical cases of minimal and conformal coupling. Comparing the leading term of $<\phi^{2}>$, proportional to the coincident limit of the Hadamard-DeWitt coeficcient $a_{2}$, with higher order terms, that include the coincident limits of coefficients up to $a_{5}$, we conclude that the next to next to next to leading approximation need to be used to give a more precise description of vacuum polarization effects in this structures. We also find the trace of the renormalized stress energy tensor for the quantized field in the leading approximation, using the existing relationship between this magnitude, the trace anomaly and the field fluctuation.
\end{abstract}
\date{\today}
\maketitle

\section{Introduction}
The General Theory of Relativity and Quantum Theory are two fundamental pillars of modern physics, which have resulted monuments of human knowledge with solidly established foundations. The General Theory of Relativity allows us to describe the cosmos, from quasars to black holes \cite{Misner}. In contrast, Quantum Theory describes the phenomena of the microworld, from atoms to quarks. Both theories have, in their scale of application, the highest reputation, thanks to their enormous predictive power.

Unfortunately, when you try to apply together to the description of the microworld, serious problems arise. All attempts to develop a conceptual framework where the General Theory of Relativity and Quantum Theory are part of a unified theory, have failed so far, so the search for a quantum theory of gravity is currently an exciting subject of enormous scientific interest.

However, at the beginning of quantum theory, some of the typical problems addressed included the interaction of microscopic systems, such as atoms and electrons, with external physical fields, such as electric and magnetic fields. In the absence of a complete quantum theory of fields, these problems were addressed considering these external fields as completely classical, while the atoms were described using quantum mechanics. This semiclassical approach allowed to obtain interesting results on the behavior of matter at the microscopic level in the presence of external fields. We can mention the basic descriptions of the Zeeman and the Stark effects as relevant examples.

Therefore, it is reasonable to think that there may be a similar regime, in which the influence of gravitation on microscopic phenomena could be analyzed. Such a semi-classical approach, initiated in Hawking's pioneering work on the radiation of black holes, is known as Quantum Field Theory in curved spacetimes, and is analogous to that used in the early years of quantum theory to address the problems mentioned above. In this approach, all matter fields, except the gravitational one, are described using quantum field theory. The gravitational field, on the other hand, is described using the general theory of relativity \cite{Hawking,DeWitt,birrel}.

In Quantum field theory in curved space-times, also called semiclassical gravity, the physical magnitudes of greatest interest to be determined are the quantum stress-energy tensor of the quantized fields, $<T_{\mu \nu}>$ and the expectation value of the square of the quantum field, or vacuum polarization $<\phi^{2}>$, where $\phi$ is the quantum field amplitude. The vacuum polarization $<\phi^{2}>$ provides information about spontaneous symmetry breaking in a given background spacetime, and can be used to determine the trace of $<T_{\mu \nu}>$ for conformally coupled scalar fields. Also, being often computed with much less effort than the stress energy tensor operator, also gives qualitative information about this quantity. The stress energy tensor of the quantum field enters as a source in the semiclassical Einstein equations that can be solved providing insight into the ways in which quantum fields can affect the spacetime background \cite{birrel,york,lousto-sanchez}.

However, the exact determination of $<\phi^{2}>$ and $<T_{\mu \nu}>$ in the general case is a very difficult task, and consequently there exist in the literature several approaches to obtain this quantity, including numerical ones \cite{candelas,koffman,frolov-zelnikov,avramidi,AHS,matyjasek1,matyjasek,berej-matyjasek,owen1,owen2,Folacci,FT}. An important case in which there exist an analytic approximation to $<\phi^{2}>$ and $<T_{\mu \nu}>$ is the related with massive fields. This approach, based on the Schwinger-DeWitt expansion of the Green\'s function of the dynamical operator that describes the evolution of the quantum field, gives good approximation whenever the Compton's wavelenght of the field is less than the characteristic radius of curvature \cite{DeWitt,frolov-zelnikov,avramidi,AHS,matyjasek1,matyjasek,berej-matyjasek,popov,owen1,owen2,Folacci}.

In the Schwinger-DeWitt approach, to determine $<T_{\mu \nu}>$ we need to functionally differentiate the effective action with respect to the metric tensor, wich is a very difficult task. However, the calculation of the vacuum polarization only requires the use the Green function.

In this paper we will focus on the determination of the vacuum polarization of a quantized  massive scalar field, with arbitrary coupling to the background gravitational field of a pointlike global monopole, whose line element was discovered by Barriola and Vilenkin \cite{barriola-vilenkin}. Monopoles are interesting systems to be considered because they constitute heavy objects formed in the early universe as a result of a phase transition of a system consisting of a self-coupled scalar field triplet whose original global $O(3)$ symmetry is spontaneously broken to $U(1)$. The scalar field plays the role of order parameter which acquires a nonvanishing value outside the monopole core, in which the main part of the monopole's energy is concentrated.

This is a first step in the investigation of vacuum polarization effects in this spacetime using the Schwinger-DeWitt approach. Previous works devoted to the quantization of quantum fields global monopole spacetime includes the analysis of massless scalar fields \cite{hiscock,mazzitely-lousto,bezerra1}, and the determination of the quantum stress energy tensor for a massless spinor field \cite{bezerra2,bezerra3}.

The paper is organized as follows. In Section II we give a brief description of the spacetime that describes a pointlike global monopole, presenting the line element that will be used as the background to find the vacuum polarization. Section III is devoted to the construction of $\left<\phi^{2}\right>$ for a massive scalar field in a general gravitational spacetime in terms of the Hadamard-DeWitt coeficients up to  the next to next to next to leading term $a_{5}$ in four dimensions. Explicit analytic results for the vacuum polarization in the spacetime of a pointlike global monopoles are presented and discussed in Section IV, whereas Section V contains our concluding remarks and some perspective about future works on this subject.

\section{The background}
Barriola and Vilenkin analysed  a simple model which gives
rise to a global monopole. The Langrangian density for the model is \cite{barriola-vilenkin}
\begin{equation}
L= {1\over 2} (\partial_\mu \phi^a)( \partial^\mu \phi^a ) - {1 \over
4} \lambda (\phi^a \phi^a - \eta^2)^2\ ,
\end{equation}
where, for a typical grand unified theory the parameter $\eta$ is of order $10^{16}Gev$. Considering the equations derived from the above Lagrangian density and the Einstein equations, we obtain the spherically symmetric solution:
\begin{equation}
ds^2 = -A(r)dt^2 + A^{-1}(r)dr^2 + r^2(d\theta^2 + \sin^2\theta
d\varphi^2)\ ,
\end{equation}
where $A(r)$ is given by
\begin{equation}
A(r) = 1 - 8\pi \eta^2 - 2 M/r\ .
\end{equation}
far from the monopole's core, and $M$ is the mass parameter. Neglecting this mass term we obtain the metric of the pointlike global monopole,
\begin{equation}
ds^2 = -\alpha^2 dt^2 + dr^2/\alpha^2 + r^2(d\theta^2 +
\sin^2\theta d\varphi^2)\ ,
\label{metric}
\end{equation}
where the parameter $\alpha^2 = 1 - 8\pi \eta^2$.

If we re-scale the time and radial variable in the above metric as $\tau=\alpha t$ and $\rho=\frac{r}{\alpha}$ we obtain the line element
\begin{equation}
ds^2 = -d\tau^2 + d\rho^2 + \alpha^{2}\rho^2(d\theta^2 +
\sin^2\theta d\varphi^2)\ ,
\label{metric2}
\end{equation}

This particular form of the line element for the global monopole, allow us to see a peculiarity of this spacetime, that  is a solid angle deficit, which is the difference between the solid angle
in the flat spacetime $4\pi$ and the solid angle in the
global monopole spacetime which is given by $4\pi\alpha^2$. When the parameter $\alpha < 1$ imply a solid angle deficit whereas $\alpha >1$ imply solid angle excess. As the physical value for $\alpha$ predicted by field theory is
smaller than unit, then we have a solid angle deficit for the global monopole spacetime.

An interesting fact derived from the line element (\ref{metric}) is that the geometry around the global monopole has non-vanishing curvature, whereas this heavy object has no Newtonian gravitational
potential, and consequently exerts no gravitational force on the matter around it, apart from the tiny gravitational effect due to the core. This gives a divergent mass for the monopole, proportional to the distance from monopole origin \cite{barriola-vilenkin}.

In \cite{HiscockPRL}, the fact that, although the global monopole has no Newtonian
gravitational potential, it gives enormous tidal acceleration
$a\sim 1/r^2$, was considered in the cosmological context to obtain an upper bound on the number
density of them in the Universe, which is at most one
global monopole in the local group galaxies. However, in \cite{BennetRhie}, was showed using numerical simulations that the real upper boundary is smaller by many orders than that derived by Hiscock in \cite{HiscockPRL}, as they find scaling solution with a few global monopoles per horizon volume.
\section{$\phi^{2}$ constructed using Schwinger-DeWitt expansion}

In this section we construct the vacuum polarization $\phi^{2}$ for a massive scalar field with mass $\mu$ and arbitrary coupling to an arbitrary gravitational background with metric tensor $g_{\mu \nu}$, using the Schwinger-DeWitt approach. We mainly follow the simple derivation given by Matyjasek and Tryniecki in reference \cite{matyjasek-tri1}.
The massive scalar field satisfy the Klein-Gordon equation
\begin{equation}
 \left(-\Box + \mu^{2} + \xi R \right) \phi = 0,
\label{kg}
\end{equation}
where $\xi$ is the coupling constant and $R$ is the Ricci scalar.

The vacuum polarization in the background spacetime is defined as \cite{hiscock}
\begin{equation}
 \langle \phi^{2} \rangle = - i \lim_{x' \to x}  G^{F}(x,x').
 \label{vacpol}
\end{equation}
where $G^{F}(x,x')$ is the Feynmann Green's of equation (\ref{kg}). As this object diverges when $x' \to x$, we need to regularize it. For this reason, in the following we use the Schwinger-DeWitt proper-time formalism which assumes that $G(x,x')$
is given by
\begin{equation}
G^{F}(x,x') =  \frac{i \Delta^{1/2}}{(4 \pi)^{2}} \int_{0}^{\infty} i ds
\frac{1}{(is)^{2}} \exp\left[-i \mu^{2} s + \frac{i \sigma(x,x')}{2 s} \right]
A(x,x'; is),
\label{grf}
\end{equation}
where
\begin{equation}
 A(x,x'; is) = \sum_{k=0}^{\infty} (is)^k a_{k}(x,x'),
\end{equation}
$s$ is the proper time and the biscalars $a_{k}(x,x')$ are called Hadamard-DeWitt
coefficients. Also $\Delta(x,x')$ is the Van-Vleck-Morette determinant and the biscalar $\sigma(x,x')$
represent one-half of the geodetic distance between the spacetime points $x$ and $x'.$

Defining the regularized biscalar $A_{reg}(x,x'; is)$ as
\begin{equation}
A_{reg}(x,x'; is) = A(x,x'; is) - \sum_{k=0}^{1}
a_{k}(x,x') (is)^k,
\end{equation}
we can put, in Eq. (\ref{grf}), $A_{reg}(x,x';is)$ instead of $A(x,x'; is)$, which gives finally the regularised four dimensional Green's function $G_{reg}(x,x')$ as:
\begin{equation}
G_{reg}(x,x')=\frac{i \Delta^{1/2}}{(4 \pi)^{2}} \int_{0}^{\infty} i ds
\frac{1}{(is)^{2}} \exp\left[-i \mu^{2} s + \frac{i \sigma(x,x')}{2 s} \right]
\sum_{k=2}^{N}
a_{k}(x,x') (is)^k,
\label{gfreg}
\end{equation}

Calculating the integral (\ref{gfreg}) by making the substitution
$\mu^{2} \to \mu^{2} - i \varepsilon$ ($\varepsilon >0$)~\cite{matyjasek-tri1}, and putting the obtained result in (\ref{vacpol}) we have the regularised field fluctuation $\langle \phi^{2} \rangle^{N}_{reg}$:
\begin{equation}
 \langle \phi^{2} \rangle^{N}_{reg} = \frac{1}{(4 \pi)^{2}} \sum_{k=2}^{N} \frac{a_{k}}{(\mu^{2})^{k-1}}  (k-2)!,
\label{main}
\end{equation}
where $a_{k} = \lim_{x' \to x} a_{k}(x,x')$ are the coincidence limits of the Hadamard-DeWitt biscalars and the upper sum limit, $N$, gives the order of the Schwinger-DeWitt approximation in $\langle \phi^{2} \rangle^{N}_{reg}$. We expect that if the Compton length associated with the field $\lambda_{c},$ is less than the characteristic radius of the curvature of the background geometry, $L$, then he leading term in (\ref{main}), proportional to the inverse of the squared field's mass, gives a reasonable approximation to $\langle \phi^{2}\rangle$.

The inclusion of higher terms in the above expansion will be always well motivated, in order to obtain a value for the field fluctuation closely to the exact value of this quantity. In the rest of the paper our aim is to take into account in (\ref{main}) terms up to the next to next to next to leading order, which imply the calculation of the coincidence limit of the Hadamard-DeWitt coefficients up to $\left[a_{5}\right]$.

\subsection{Hadamard-DeWitt coefficients}

As we see from (\ref{main}), the main task for the calculation of the vacuum polarization is the determination of the coincidence limit of the Hadamard-DeWitt biscalars $a_{k}(x,x')$, that satisfy the recurrence equation
\begin{equation}
\sigma^{;i} a_{k;i} + k a_{k} - \Delta^{-1/2}\Box \left( \Delta^{1/2}
a_{k-1}\right) + \xi R a_{k-1} =0,
\label{DeWitt}
\end{equation}
with the boundary condition $a_{0}(x,x')=1.$

Up to now, there are some previous works that give this coefficients up to order $k=5$ \cite{DeWitt,avramidi,gilkey,wardel,matyjasek4}. In this work, we used the transport equation approach of Ottewill and Wardell \cite{wardel}, and obtained general expressions for  the coincidence limit of the Hadamard-DeWitt coefficients up to $a_{5}$, by solving the transport equations given in \cite{wardel} using the software package xAct for Wolfram Mathematica \cite{xact}. Then we have the results $a_{0}=1$, $a_{1}=\frac{1}{6} R$, and, up to $N=5$:
\begin{IEEEeqnarray}{rClrClrCl}
a_{2} &=& \frac{1}{360} (-2 R{}_{\alpha }{}_{\beta } R{}^{\alpha }{}^{\beta } + 5 R^{2} + 2 R{}_{\alpha }{}_{\beta }{}_{\gamma }{}_{\delta } R{}^{\alpha }{}^{\beta }{}^{\gamma }{}^{\delta } + 12 R{}^{;\alpha }{}_{\alpha })
\end{IEEEeqnarray}
\begin{IEEEeqnarray}{rCl}
a_{3} &=& \frac{1}{15120} (584 R{}_{\alpha }{}^{\gamma } R{}^{\alpha }{}^{\beta } R{}_{\beta }{}_{\gamma } - 654 R{}_{\alpha }{}_{\beta } R{}^{\alpha }{}^{\beta } R + 99 R^{3} + 456 R{}^{\alpha }{}^{\beta } R{}^{\gamma }{}^{\delta } R{}_{\alpha }{}_{\gamma }{}_{\beta }{}_{\delta } + 72 R R{}_{\alpha }{}_{\beta }{}_{\gamma }{}_{\delta } R{}^{\alpha }{}^{\beta }{}^{\gamma }{}^{\delta } \nonumber \\
&& - 80 R{}_{\alpha }{}_{\beta }{}^{\epsilon }{}^{\rho } R{}^{\alpha }{}^{\beta }{}^{\gamma }{}^{\delta } R{}_{\gamma }{}_{\delta }{}_{\epsilon }{}_{\rho } + 51 R{}_{;\alpha } R{}^{;\alpha } - 12 R{}_{\alpha }{}_{\gamma }{}_{;\beta } R{}^{\alpha }{}^{\beta }{}^{;\gamma } - 6 R{}_{\alpha }{}_{\beta }{}_{;\gamma } R{}^{\alpha }{}^{\beta }{}^{;\gamma } + 27 R{}_{\alpha }{}_{\beta }{}_{\gamma }{}_{\delta }{}_{;\epsilon } R{}^{\alpha }{}^{\beta }{}^{\gamma }{}^{\delta }{}^{;\epsilon } \nonumber \\
&& + 84 R R{}^{;\alpha }{}_{\alpha } + 36 R{}_{\alpha }{}_{\beta } R{}^{;\alpha }{}^{\beta } - 24 R{}^{\alpha }{}^{\beta } R{}_{\alpha }{}_{\beta }{}^{;\gamma }{}_{\gamma } + 144 R{}_{\alpha }{}_{\gamma }{}_{\beta }{}_{\delta } R{}^{\alpha }{}^{\beta }{}^{;\gamma }{}^{\delta } + 54 R{}^{;\alpha }{}_{\alpha }{}^{\beta }{}_{\beta })\nonumber \\
&&+ \frac{\xi}{360} (2 R{}_{\alpha }{}_{\beta } R{}^{\alpha }{}^{\beta } R - 5 R^{3} - 2 R R{}_{\alpha }{}_{\beta }{}_{\gamma }{}_{\delta } R{}^{\alpha }{}^{\beta }{}^{\gamma }{}^{\delta } - 12 R{}_{;\alpha } R{}^{;\alpha } - 22 R R{}^{;\alpha }{}_{\alpha } - 4 R{}_{\alpha }{}_{\beta } R{}^{;\alpha }{}^{\beta } - 6 R{}^{;\alpha }{}_{\alpha }{}^{\beta }{}_{\beta })\nonumber \\
&&+\frac{\xi^{2}}{12} (R^{3} + R{}_{;\alpha } R{}^{;\alpha } + 2 R R{}^{;\alpha }{}_{\alpha })-\frac{\xi^{3}}{6} R^{3} 
\end{IEEEeqnarray}
\begin{IEEEeqnarray}{rCl}
a_{4}&=& \frac{1}{1814400} (-32736 R{}_{\alpha }{}^{\gamma } R{}^{\alpha }{}^{\beta } R{}_{\beta }{}^{\delta } R{}_{\gamma }{}_{\delta } + 8436 R{}_{\alpha }{}_{\beta } R{}^{\alpha }{}^{\beta } R{}_{\gamma }{}_{\delta } R{}^{\gamma }{}^{\delta } + 59136 R{}_{\alpha }{}^{\gamma } R{}^{\alpha }{}^{\beta } R{}_{\beta }{}_{\gamma } R  \nonumber \\
&& - 43518 R{}_{\alpha }{}_{\beta } R{}^{\alpha }{}^{\beta } R^{2}+ 2700 R R{}_{\alpha }{}_{\beta }{}_{\gamma }{}_{\delta }{}_{;\epsilon } R{}^{\alpha }{}^{\beta }{}^{\gamma }{}^{\delta }{}^{;\epsilon } + 8352 R R{}_{\alpha }{}_{\gamma }{}_{\beta }{}_{\delta } R{}^{\alpha }{}^{\beta }{}^{;\gamma }{}^{\delta }\nonumber\\
&& + 5743 R^{4} + 13944 R{}^{\alpha }{}^{\beta } R{}^{\gamma }{}^{\delta } R R{}_{\alpha }{}_{\gamma }{}_{\beta }{}_{\delta } + 3618 R^{2} R{}_{\alpha }{}_{\beta }{}_{\gamma }{}_{\delta } R{}^{\alpha }{}^{\beta }{}^{\gamma }{}^{\delta } + 168 R{}^{\alpha }{}^{\beta } R{}^{\gamma }{}^{\delta } R{}_{\alpha }{}_{\gamma }{}^{\epsilon }{}^{\rho } R{}_{\beta }{}_{\delta }{}_{\epsilon }{}_{\rho } \nonumber \\
&& + 14832 R{}^{\alpha }{}^{\beta } R{}^{\gamma }{}^{\delta } R{}_{\alpha }{}^{\epsilon }{}_{\beta }{}^{\rho } R{}_{\gamma }{}_{\epsilon }{}_{\delta }{}_{\rho } - 3282 R{}_{\alpha }{}_{\beta } R{}^{\alpha }{}^{\beta } R{}_{\gamma }{}_{\delta }{}_{\epsilon }{}_{\rho } R{}^{\gamma }{}^{\delta }{}^{\epsilon }{}^{\rho } - 2496 R{}_{\alpha }{}_{\beta }{}^{\epsilon }{}^{\rho } R{}^{\alpha }{}^{\beta }{}^{\gamma }{}^{\delta } R{}_{\gamma }{}_{\epsilon }{}^{\sigma }{}^{\tau } R{}_{\delta }{}_{\rho }{}_{\sigma }{}_{\tau } \nonumber \\
&&  - 4480 R R{}_{\alpha }{}_{\beta }{}^{\epsilon }{}^{\rho } R{}^{\alpha }{}^{\beta }{}^{\gamma }{}^{\delta } R{}_{\gamma }{}_{\delta }{}_{\epsilon }{}_{\rho }+ 1248 R{}_{\alpha }{}_{\beta }{}^{\epsilon }{}^{\rho } R{}^{\alpha }{}^{\beta }{}^{\gamma }{}^{\delta } R{}_{\gamma }{}_{\delta }{}^{\sigma }{}^{\tau } R{}_{\epsilon }{}_{\rho }{}_{\sigma }{}_{\tau } + 696 R{}_{\alpha }{}_{\beta }{}_{\gamma }{}_{\delta } R{}^{\alpha }{}^{\beta }{}^{\gamma }{}^{\delta } R{}_{\epsilon }{}_{\rho }{}_{\sigma }{}_{\tau } R{}^{\epsilon }{}^{\rho }{}^{\sigma }{}^{\tau } \nonumber \\
&&  - 65040 R{}^{\beta }{}^{\gamma } R{}_{\beta }{}_{\gamma }{}_{;\alpha } R{}^{;\alpha } + 13740 R R{}_{;\alpha } R{}^{;\alpha }- 2160 R R{}_{\alpha }{}_{\gamma }{}_{;\beta } R{}^{\alpha }{}^{\beta }{}^{;\gamma } - 11136 R{}^{\alpha }{}^{\beta } R{}^{\gamma }{}^{\delta } R{}_{\alpha }{}_{\gamma }{}_{;\beta }{}_{\delta }\nonumber\\
&& + 6960 R{}^{\beta }{}^{\gamma }{}^{\delta }{}^{\epsilon } R{}_{\beta }{}_{\gamma }{}_{\delta }{}_{\epsilon }{}_{;\alpha } R{}^{;\alpha } + 1440 R{}^{\alpha }{}^{\beta } R{}^{\gamma }{}^{\delta }{}_{;\alpha } R{}_{\gamma }{}_{\delta }{}_{;\beta } - 1560 R{}_{\alpha }{}_{\beta } R{}^{;\alpha } R{}^{;\beta } + 2880 R{}^{\beta }{}^{\gamma } R{}^{;\alpha } R{}_{\alpha }{}_{\beta }{}_{;\gamma } \nonumber \\
&& - 5760 R{}_{\alpha }{}^{\delta }{}^{\epsilon }{}^{\rho } R{}_{\gamma }{}_{\delta }{}_{\epsilon }{}_{\rho }{}_{;\beta } R{}^{\alpha }{}^{\beta }{}^{;\gamma } - 28920 R R{}_{\alpha }{}_{\beta }{}_{;\gamma } R{}^{\alpha }{}^{\beta }{}^{;\gamma } + 27840 R{}_{\alpha }{}_{\delta }{}_{\beta }{}_{\epsilon } R{}^{\delta }{}^{\epsilon }{}_{;\gamma } R{}^{\alpha }{}^{\beta }{}^{;\gamma } - 7680 R{}^{\alpha }{}^{\beta } R{}_{\gamma }{}_{\delta }{}_{;\beta } R{}_{\alpha }{}^{\gamma }{}^{;\delta } \nonumber \\
&& + 4800 R{}^{\alpha }{}^{\beta } R{}_{\beta }{}_{\delta }{}_{;\gamma } R{}_{\alpha }{}^{\gamma }{}^{;\delta } + 85440 R{}^{\alpha }{}^{\beta } R{}_{\beta }{}_{\gamma }{}_{;\delta } R{}_{\alpha }{}^{\gamma }{}^{;\delta } - 1920 R{}_{\alpha }{}_{\beta }{}_{\gamma }{}_{\delta } R{}^{;\alpha } R{}^{\beta }{}^{\gamma }{}^{;\delta } - 1920 R{}_{\beta }{}_{\gamma }{}_{\delta }{}_{\epsilon } R{}^{\alpha }{}^{\beta }{}^{;\gamma } R{}_{\alpha }{}^{\delta }{}^{;\epsilon } \nonumber \\
&& - 7680 R{}_{\alpha }{}_{\delta }{}_{\beta }{}_{\epsilon } R{}^{\alpha }{}^{\beta }{}^{;\gamma } R{}_{\gamma }{}^{\delta }{}^{;\epsilon } + 14400 R{}^{\alpha }{}^{\beta } R{}_{\alpha }{}_{\gamma }{}_{\beta }{}_{\epsilon }{}_{;\delta } R{}^{\gamma }{}^{\delta }{}^{;\epsilon } + 34080 R{}^{\alpha }{}^{\beta } R{}_{\alpha }{}_{\gamma }{}_{\beta }{}_{\delta }{}_{;\epsilon } R{}^{\gamma }{}^{\delta }{}^{;\epsilon }  \nonumber \\
&& + 12960 R{}^{\alpha }{}^{\beta } R{}_{\beta }{}_{\delta }{}_{\gamma }{}_{\rho }{}_{;\epsilon } R{}_{\alpha }{}^{\gamma }{}^{\delta }{}^{\epsilon }{}^{;\rho } - 10800 R{}^{\alpha }{}^{\beta }{}^{\gamma }{}^{\delta } R{}_{\gamma }{}_{\delta }{}_{\epsilon }{}_{\rho }{}_{;\sigma } R{}_{\alpha }{}_{\beta }{}^{\epsilon }{}^{\rho }{}^{;\sigma } - 9792 R{}_{\beta }{}_{\gamma } R{}^{\beta }{}^{\gamma } R{}^{;\alpha }{}_{\alpha } + 4608 R^{2} R{}^{;\alpha }{}_{\alpha } \nonumber \\
&& + 1296 R{}_{\beta }{}_{\gamma }{}_{\delta }{}_{\epsilon } R{}^{\beta }{}^{\gamma }{}^{\delta }{}^{\epsilon } R{}^{;\alpha }{}_{\alpha } + 432 R{}_{\alpha }{}_{\beta } R R{}^{;\alpha }{}^{\beta } + 1632 R{}^{\gamma }{}^{\delta } R{}_{\alpha }{}_{\gamma }{}_{\beta }{}_{\delta } R{}^{;\alpha }{}^{\beta } + 936 R{}_{;\alpha }{}_{\beta } R{}^{;\alpha }{}^{\beta }  \nonumber \\
&& + 1008 R{}^{;\alpha }{}_{\alpha } R{}^{;\beta }{}_{\beta } + 10464 R{}^{\alpha }{}^{\beta } R{}^{\gamma }{}^{\delta } R{}_{\alpha }{}_{\beta }{}_{;\gamma }{}_{\delta } - 18000 R{}^{\alpha }{}^{\beta } R R{}_{\alpha }{}_{\beta }{}^{;\gamma }{}_{\gamma } + 624 R{}^{;\alpha }{}^{\beta } R{}_{\alpha }{}_{\beta }{}^{;\gamma }{}_{\gamma } \nonumber \\
&& - 14016 R{}_{\alpha }{}_{\gamma }{}^{\epsilon }{}^{\rho } R{}_{\beta }{}_{\delta }{}_{\epsilon }{}_{\rho } R{}^{\alpha }{}^{\beta }{}^{;\gamma }{}^{\delta } + 384 R{}_{\alpha }{}^{\epsilon }{}_{\beta }{}^{\rho } R{}_{\gamma }{}_{\epsilon }{}_{\delta }{}_{\rho } R{}^{\alpha }{}^{\beta }{}^{;\gamma }{}^{\delta } + 1872 R{}_{\gamma }{}_{\delta }{}_{;\alpha }{}_{\beta } R{}^{\alpha }{}^{\beta }{}^{;\gamma }{}^{\delta } - 4032 R{}_{\alpha }{}_{\gamma }{}_{;\beta }{}_{\delta } R{}^{\alpha }{}^{\beta }{}^{;\gamma }{}^{\delta } \nonumber \\
&& + 1872 R{}_{\alpha }{}_{\beta }{}_{;\gamma }{}_{\delta } R{}^{\alpha }{}^{\beta }{}^{;\gamma }{}^{\delta } + 2304 R{}^{\alpha }{}^{\beta } R{}^{\gamma }{}^{\delta }{}^{\epsilon }{}^{\rho } R{}_{\alpha }{}_{\gamma }{}_{\beta }{}_{\epsilon }{}_{;\delta }{}_{\rho } - 216 R{}^{\alpha }{}^{\beta }{}^{;\gamma }{}_{\gamma } R{}_{\alpha }{}_{\beta }{}^{;\delta }{}_{\delta } + 23904 R{}_{\alpha }{}^{\gamma } R{}^{\alpha }{}^{\beta } R{}_{\beta }{}_{\gamma }{}^{;\delta }{}_{\delta } \nonumber \\
&& + 8448 R{}^{\alpha }{}^{\beta } R{}_{\beta }{}_{\delta }{}_{\gamma }{}_{\epsilon } R{}_{\alpha }{}^{\gamma }{}^{;\delta }{}^{\epsilon } + 12288 R{}^{\alpha }{}^{\beta } R{}_{\alpha }{}_{\gamma }{}_{\beta }{}_{\delta } R{}^{\gamma }{}^{\delta }{}^{;\epsilon }{}_{\epsilon } + 576 R{}_{\alpha }{}_{\beta }{}_{\gamma }{}_{\delta }{}_{;\epsilon }{}_{\rho } R{}^{\alpha }{}^{\beta }{}^{\gamma }{}^{\delta }{}^{;\epsilon }{}^{\rho } + 2640 R{}^{;\alpha } R{}_{;\alpha }{}^{\beta }{}_{\beta } \nonumber \\
&& + 960 R{}_{\alpha }{}_{\beta }{}_{;\gamma } R{}^{;\alpha }{}^{\beta }{}^{\gamma } - 480 R{}^{\alpha }{}^{\beta }{}^{;\gamma } R{}_{\alpha }{}_{\gamma }{}_{;\beta }{}^{\delta }{}_{\delta } - 240 R{}^{\alpha }{}^{\beta }{}^{;\gamma } R{}_{\alpha }{}_{\beta }{}_{;\gamma }{}^{\delta }{}_{\delta } + 5760 R{}_{\alpha }{}_{\gamma }{}_{\beta }{}_{\delta }{}_{;\epsilon } R{}^{\alpha }{}^{\beta }{}^{;\gamma }{}^{\delta }{}^{\epsilon } \nonumber \\
&& + 1080 R R{}^{;\alpha }{}_{\alpha }{}^{\beta }{}_{\beta } + 960 R{}_{\alpha }{}_{\beta } R{}^{;\alpha }{}^{\beta }{}^{\gamma }{}_{\gamma } - 240 R{}^{\alpha }{}^{\beta } R{}_{\alpha }{}_{\beta }{}^{;\gamma }{}_{\gamma }{}^{\delta }{}_{\delta } + 1920 R{}_{\alpha }{}_{\gamma }{}_{\beta }{}_{\delta } R{}^{\alpha }{}^{\beta }{}^{;\gamma }{}^{\delta }{}^{\epsilon }{}_{\epsilon } + 480 R{}^{;\alpha }{}_{\alpha }{}^{\beta }{}_{\beta }{}^{\gamma }{}_{\gamma }) \nonumber \\
&& +\frac{\xi}{15120} (-584 R{}_{\alpha }{}^{\gamma } R{}^{\alpha }{}^{\beta } R{}_{\beta }{}_{\gamma } R + 654 R{}_{\alpha }{}_{\beta } R{}^{\alpha }{}^{\beta } R^{2} - 99 R^{4} - 456 R{}^{\alpha }{}^{\beta } R{}^{\gamma }{}^{\delta } R R{}_{\alpha }{}_{\gamma }{}_{\beta }{}_{\delta } - 72 R^{2} R{}_{\alpha }{}_{\beta }{}_{\gamma }{}_{\delta } R{}^{\alpha }{}^{\beta }{}^{\gamma }{}^{\delta } \nonumber \\
&& + 80 R R{}_{\alpha }{}_{\beta }{}^{\epsilon }{}^{\rho } R{}^{\alpha }{}^{\beta }{}^{\gamma }{}^{\delta } R{}_{\gamma }{}_{\delta }{}_{\epsilon }{}_{\rho } + 12 R{}^{\beta }{}^{\gamma } R{}_{\beta }{}_{\gamma }{}_{;\alpha } R{}^{;\alpha } - 135 R R{}_{;\alpha } R{}^{;\alpha } - 36 R{}^{\beta }{}^{\gamma }{}^{\delta }{}^{\epsilon } R{}_{\beta }{}_{\gamma }{}_{\delta }{}_{\epsilon }{}_{;\alpha } R{}^{;\alpha } + 102 R{}_{\alpha }{}_{\beta } R{}^{;\alpha } R{}^{;\beta } \nonumber \\
&& + 24 R{}^{\beta }{}^{\gamma } R{}^{;\alpha } R{}_{\alpha }{}_{\beta }{}_{;\gamma } + 12 R R{}_{\alpha }{}_{\gamma }{}_{;\beta } R{}^{\alpha }{}^{\beta }{}^{;\gamma } + 6 R R{}_{\alpha }{}_{\beta }{}_{;\gamma } R{}^{\alpha }{}^{\beta }{}^{;\gamma } + 24 R{}_{\alpha }{}_{\beta }{}_{\gamma }{}_{\delta } R{}^{;\alpha } R{}^{\beta }{}^{\gamma }{}^{;\delta } - 27 R R{}_{\alpha }{}_{\beta }{}_{\gamma }{}_{\delta }{}_{;\epsilon } R{}^{\alpha }{}^{\beta }{}^{\gamma }{}^{\delta }{}^{;\epsilon } \nonumber \\
&& + 30 R{}_{\beta }{}_{\gamma } R{}^{\beta }{}^{\gamma } R{}^{;\alpha }{}_{\alpha } - 123 R^{2} R{}^{;\alpha }{}_{\alpha } - 18 R{}_{\beta }{}_{\gamma }{}_{\delta }{}_{\epsilon } R{}^{\beta }{}^{\gamma }{}^{\delta }{}^{\epsilon } R{}^{;\alpha }{}_{\alpha } - 48 R{}_{\alpha }{}_{\beta } R R{}^{;\alpha }{}^{\beta } - 72 R{}^{\gamma }{}^{\delta } R{}_{\alpha }{}_{\gamma }{}_{\beta }{}_{\delta } R{}^{;\alpha }{}^{\beta } \nonumber \\
&& - 72 R{}_{;\alpha }{}_{\beta } R{}^{;\alpha }{}^{\beta } - 84 R{}^{;\alpha }{}_{\alpha } R{}^{;\beta }{}_{\beta } + 24 R{}^{\alpha }{}^{\beta } R R{}_{\alpha }{}_{\beta }{}^{;\gamma }{}_{\gamma } - 24 R{}^{;\alpha }{}^{\beta } R{}_{\alpha }{}_{\beta }{}^{;\gamma }{}_{\gamma } - 144 R R{}_{\alpha }{}_{\gamma }{}_{\beta }{}_{\delta } R{}^{\alpha }{}^{\beta }{}^{;\gamma }{}^{\delta } \nonumber \\
&& - 210 R{}^{;\alpha } R{}_{;\alpha }{}^{\beta }{}_{\beta } - 36 R{}_{\alpha }{}_{\beta }{}_{;\gamma } R{}^{;\alpha }{}^{\beta }{}^{\gamma } - 96 R R{}^{;\alpha }{}_{\alpha }{}^{\beta }{}_{\beta } - 36 R{}_{\alpha }{}_{\beta } R{}^{;\alpha }{}^{\beta }{}^{\gamma }{}_{\gamma } - 18 R{}^{;\alpha }{}_{\alpha }{}^{\beta }{}_{\beta }{}^{\gamma }{}_{\gamma }) \nonumber \\
&& +\frac{\xi^{2}}{720} (-2 R{}_{\alpha }{}_{\beta } R{}^{\alpha }{}^{\beta } R^{2} + 5 R^{4} + 2 R^{2} R{}_{\alpha }{}_{\beta }{}_{\gamma }{}_{\delta } R{}^{\alpha }{}^{\beta }{}^{\gamma }{}^{\delta } + 34 R R{}_{;\alpha } R{}^{;\alpha } - 12 R{}_{\alpha }{}_{\beta } R{}^{;\alpha } R{}^{;\beta } + 32 R^{2} R{}^{;\alpha }{}_{\alpha } \nonumber \\
&& + 8 R{}_{\alpha }{}_{\beta } R R{}^{;\alpha }{}^{\beta } + 8 R{}_{;\alpha }{}_{\beta } R{}^{;\alpha }{}^{\beta } + 10 R{}^{;\alpha }{}_{\alpha } R{}^{;\beta }{}_{\beta } + 24 R{}^{;\alpha } R{}_{;\alpha }{}^{\beta }{}_{\beta } + 12 R R{}^{;\alpha }{}_{\alpha }{}^{\beta }{}_{\beta }) \nonumber \\
&& +\frac{\xi^{3}}{36} (- R^{4} - 3 R R{}_{;\alpha } R{}^{;\alpha } - 3 R^{2} R{}^{;\alpha }{}_{\alpha })+ \frac{\xi^{4}}{24} R^{4} 
\end{IEEEeqnarray}


We checked that the above general results coincide with those reported in references \cite{wardel,matyjasek4}. As we can see, Hadamard-DeWitt coefficients are extremely complicated local expressions constructed from the Riemmann tensor, their covariant derivatives, and contractions. However, the fact that the above results are valid for a generic spacetime, being static, stationary or non-stationary, gives rise to the possibility of use them to obtain relatively simple expressions for the field fluctuation of quantum massive scalar field in spacetimes with higher degree of symmetry.

As we can see from the structure of the coincidence limits of Hadamard-DeWitt coefficients above, they are local geometric terms that depends of the coupling constant $\xi$ and the parameters that describe the geometry of the gravitational background. For the pointlike global monopole spacetime, the only parameter that characterizes the geometry of the manifold is that of which depends the angle deficit factor, i.e, $\alpha$. Then, we expect that the vacuum polarization let be a function of $\xi$, $\alpha$ and the distance $r$ from the monopole core.
\section{Vacuum polarization in pointlike global monopole spacetime}

\subsection{General results}

Evaluating the general formulas obtained in the spacetime background of a pointlike global monopole (\ref{metric}) we obtain simple results for the coincidence limits of the Hadamard-DeWitt coefficients:
\begin{equation}
  {a_{1}}=\,{\frac {\left(6\xi-1\right)\left({\alpha}^{2}-1\right)}{3{r}^{2
}}}
\label{a1}
\end{equation}
\begin{equation}
{a_{2}}=\,{\frac { \left( {\alpha}^{2}-1 \right)}{15{r}^{4}}}\left[ 30\,{\xi}^{2}\left({\alpha}^{2}-1\right)-{\alpha}^{2}+10\,\xi-1 \right]
\end{equation}
\begin{equation}
{a_{3}}={\frac { \left( {\alpha}^{2}-1 \right)  \left[ \beta_{1}\, {\xi}^{3}+  \beta_{2}\,{\xi}^{2}+ \beta_{3}\,\xi-25\,{\alpha}^{4}-25\,{\alpha}^{2}-4 \right] }{315\,{r}^{6
}}}
\end{equation}
where $\beta_{1}= 420\,{\alpha}^{4}-840\,{\alpha}^{2}+420$, $\beta_{2}= 630\,{\alpha}^{4}-
420\,{\alpha}^{2}-210$ and $\beta_{3}= 210\,{\alpha}^{2}+42$.
For $a_{4}$ we obtain the result
\begin{equation}
{a_{4}}={\frac { \left(1- {\alpha}^{2} \right)\left[\lambda_{0}(r,\alpha)+ \lambda_{1}(r,\alpha)\,{\xi}^{4}+\lambda_{2}(r,\alpha)\,{\xi}^{3}+ \lambda_{3}(r,\alpha)\,{\xi}^{2}+
 \lambda_{4}(r,\alpha)\, \xi\right] }{945\,{r}^{10
}}}
\end{equation}
where $\lambda_{0}(r,\alpha)=161\,{\alpha}^{6}{r}^{2}+161\,{\alpha}^{4}{r}^{2}+35\,{
\alpha}^{2}{r}^{2}+3\,{r}^{2}$, $\lambda_{1}(r,\alpha)=-630\,{\alpha}^{6}{r}^{2}+1890\,{\alpha}^{4}{r}^{2}-1890\,{
\alpha}^{2}{r}^{2}+630\,{r}^{2}$, $\lambda_{2}(r,\alpha)=-3360\,{
\alpha}^{6}{r}^{2}+6300\,{\alpha}^{4}{r}^{2}-2520\,{\alpha}^{2}{r}^{2}
-420\,{r}^{2}$, $\lambda_{3}(r,\alpha)= 3696\,{\alpha}^{8}-1932\,{
\alpha}^{6}{r}^{2}-7392\,{\alpha}^{6}+252\,{\alpha}^{4}{r}^{2}+3696\,{
\alpha}^{4}+1554\,{\alpha}^{2}{r}^{2}+126\,{r}^{2} $ and $\lambda_{4}(r,\alpha)=-1260\,{\alpha}^{4}{r}^{2}-336\,{\alpha}^{2}{r}^{2}-24\,{r}^{2}$.

The last coefficient that we considered in this work, $a_{5}$, is given by
\begin{equation}
{a_{5}}={\frac { \left(1- {\alpha}^{2} \right)\{\omega_{0}(r,\alpha)+ {\alpha}^{4}{r}^{6}\left[\omega_{1}(\alpha)\,{\xi}^{4}+\omega_{2}(\alpha)\,{\xi}^{3}+ \omega_{3}(\alpha)\,{\xi}^{2}+
 \omega_{4}(\alpha)\, \xi\right]\}}{1871100\,{r}^{16}}}
\end{equation}
where
\begin{IEEEeqnarray}{rCl}
\omega_{0}(r,\alpha) &=&\alpha^{2}\left[46860\left(1+\alpha^{8}\right)-117150\alpha^{2}\left(1+\alpha^{4}\right)+156200\alpha^{4}-7810\alpha^{10}\right] \nonumber\\
&&+4862649\alpha^{2}r^{2}\left(1+\alpha^{4}+\alpha^{6}+\alpha^{10}\right)-24313245\alpha^{4}r^{2}\left(1+\alpha^{6}\right)\nonumber\\
&&+\alpha^{4}r^{4}\left[-7461732\left(1+\alpha^{8}\right)+29846928\alpha^{2}(1+\alpha^{4})-44770392\alpha^{4}\right]\nonumber\\
&&+\alpha^{4}r^{6}\left(2160+1546549\alpha^{2}-4170957\alpha^{4}+5283387\alpha^{6}-393139\alpha^{8}\right)\nonumber\\
&&+70272\alpha^{10}r^{8}\left(1+\alpha^{2}\right)+49202\alpha^{10}r^{10}\left(\alpha^{2}-1\right),
\end{IEEEeqnarray}
\begin{equation}
\omega_{1}(\alpha)=\left( -7484400\,{\alpha}^{6}-2245320\,{\alpha}^{4}-237600\,{\alpha}
^{2}-11880 \right),
\end{equation}
\begin{equation}
\omega_{2}(\alpha)=
\left( -22453200\,{\alpha}^{12}+11226600\,{\alpha}^{10}+9979200\,{
\alpha}^{8}+1199880\,{\alpha}^{6}+47520 \right),
\end{equation}
\begin{equation}
\omega_{3}(\alpha)=\left( -22453200\,{\alpha}^{8}+40748400\,{\alpha}^{6}-14303520\,{
\alpha}^{4}-3825360\,{\alpha}^{2}-166320 \right),
\end{equation}
\begin{equation}
\omega_{4}(\alpha)= \left(-6237000\alpha^{8}+18295200\alpha^{6}-17463600\alpha^{4}+4989600\alpha^{2}+415800 \right),
\end{equation}
and
\begin{equation}
\omega_{5}(\alpha)=498960 \left( -{\alpha}^{8}+4\,{\alpha}^{6}-6\,{\alpha}^{4}+2\,{\alpha}^{2}-1 \right).
\label{a5last}
\end{equation}

The above expressions show that all the geometric coefficients $\left[a_{k}\right]$ are decreasing functions of the distance $r$ to monopole's core. Substituting (\ref{a1}) to (\ref{a5last}) in (\ref{main}), we can obtain analytic formulae for the field fluctuation $<\phi^{2}>_{reg}$ that describes vacuum polarization, for arbitrary values of the coupling parameter $\xi$, in the pointlike global monopole spacetime.

At this point we distinguish four main approximations. The leading approximation $\left <\phi^{2}\right >^{\textit{(1)}}$ is obtained evaluating the first term in (\ref{main}), that is proportional to the coefficient $a_{2}$, and to the inverse squared mass of the scalar field:
\begin{equation}
\left <\phi^{2}\right >^{\textit{(1)}}={\frac {a_{{2}}}{16{\pi}^{2}{\mu}^{2}}}
\end{equation}
For the pointlike global monopole spacetime the result is
\begin{equation}
\left <\phi^{2}\right >^{\textit{(1)}}=\frac {\left (\alpha^2 -
      1 \right)\left (30\alpha^2\xi^2 - \alpha^2 - 30\xi^2 + 10\xi -
     1 \right)} {240\pi^2\mu^2 r^4}
\end{equation}

Higher order terms includes higher Hadamard-DeWitt coefficients and higher powers of the squared inverse mass of the field. The next to leading term $\left <\phi^{2}\right >^{\textit{(2)}}$ includes $a_{2}$ and $a_{3}$, and is given by
\begin{equation}
\left <\phi^{2}\right >^{\textit{(2)}}={\frac {1}{{16}{\pi}^{2}{\mu}^{2}}}\left(a_{{2}}+\frac{a_{{3}}}{{\mu}^{2}}\right)
\end{equation}
which results for the pointlike global monopole background
\begin{equation}
\left <\phi^{2}\right >^{\textit{(2)}}=\frac {\left (\alpha^2 - 1 \right)}{5040\pi^2\mu^4 r^6}\left [\eta_{0}(r)+\eta_{1}(r)\xi+\eta_{2}(r)\xi^2+\eta_{3}(r)\xi^3\right]
\end{equation}
where $\eta_{0}(r)=-\left(25 \alpha ^2-21 \mu ^2 r^2\right)\left(\alpha^2+1\right)-4$,\ $\eta_{1}(r)=210 \left(\alpha ^2+\mu ^2 r^2\right)+42$, \ $\eta_{2}(r)=630 \left[\alpha ^4+\mu ^2 r^2\left(\alpha^{2}-1\right)\right]-210\left(2\alpha^{2}+1\right)$ \ and $\eta_{3}(r)=420\left[\alpha^2\left(\alpha^2-2\right)+1\right]$.

The next to next to leading approximation for the field fluctuation $\left <\phi^{2}\right >^{\textit{(3)}}$ includes also $a_{4}$:
\begin{equation}
\left <\phi^{2}\right >^{\textit{(3)}}={\frac {1}{{16}{\pi}^{2}{\mu}^{2}}}\left(a_{{2}}+\frac{a_{{3}}}{{\mu}^{2}}+\frac{2\,a_{{4}}}{{\mu}^{4}}\right)
\end{equation}
The result is
\begin{equation}
\left <\phi^{2}\right >^{\textit{(3)}}=\frac {\left (1-\alpha^2\right)}{15120 \pi ^2 \mu ^6 r^{10}}\sum^{3}_{j=0}\sum^{4}_{k=0}\kappa_{jk}(\alpha, \mu)r^{2j}\xi^{k}
\label{phi3}
\end{equation}
where the coefficients $\kappa_{jk}(\alpha, \mu)$ that are non zero are given by
\begin{equation}
\kappa_{10}(\alpha, \mu)=322\left( \alpha ^6+\alpha^4\right)+70\alpha ^2+6
\label{phi3coef10}
\end{equation}
\begin{equation}
\kappa_{20}(\alpha, \mu)=75\alpha^{2}\mu^{2}\left(1+\alpha ^2\right)+12 \mu ^2,
\label{phi3coef20}
\end{equation}
\begin{equation}
\kappa_{30}(\alpha, \mu)=63\mu^4\left(\alpha^{2}+1\right),
\label{phi3coef30}
\end{equation}
\begin{equation}
\kappa_{11}(\alpha, \mu)=-\alpha ^2\left(2520 \alpha ^2+672\right)-48,
\end{equation}
\begin{equation}
\kappa_{21}(\alpha, \mu)=-126\mu^{2}\left(5\alpha^{2}-1\right),
\end{equation}
\begin{equation}
\kappa_{31}(\alpha, \mu)=-630\mu^{4},
\end{equation}
\begin{equation}
\kappa_{02}(\alpha, \mu)=7392\alpha^{4}\left(1+2\alpha^{6}+\alpha^{8}\right),
\end{equation}
\begin{equation}
\kappa_{12}(\alpha, \mu)=252+\alpha^{2}\left(-3864\alpha^{3}+504\alpha^{2}+3108\right),
\end{equation}
\begin{equation}
\kappa_{22}(\alpha, \mu)=630\mu^{2}\left[1+\alpha^{2}\left(2-3\alpha^{2}\right)\right],
\end{equation}
\begin{equation}
\kappa_{32}(\alpha, \mu)=1890\mu^{4}\left(1-\alpha^{2}\right)
\end{equation}
\begin{equation}
\kappa_{13}(\alpha, \mu)=8040\left[\alpha^{2}\left(6+15\alpha^{2}-8\alpha^{4}\right)-1\right],
\end{equation}
\begin{equation}
\kappa_{23}(\alpha, \mu)=1260\mu^{2}\left[\alpha^{2}\left(2-\alpha^{2}\right)-1\right]
\end{equation}
and
\begin{equation}
\kappa_{14}(\alpha,\mu)=1260\left[1+\alpha^{2}\left(\alpha^{4}+3\alpha^{2}-3\right)\right]
\end{equation}

Finally, the next to next to next to leading term $\left <\phi^{2}\right >^{\textit{(4)}}$, that includes $\left[a_{5}\right]$ is given by
\begin{equation}
\left <\phi^{2}\right >^{\textit{(4)}}={\frac {1}{{16}{\pi}^{2}{\mu}^{2}}}\left(a_{{2}}+\frac{a_{{3}}}{{\mu}^{2}}+\frac{2\,a_{{4}}}{{\mu}^{4}}+\frac{6\,a_{{5}}}{{\mu}^{6}}\right)
\end{equation}
which in the metric of our interest results in
\begin{equation}
\left <\phi^{2}\right >^{\textit{(4)}}=\frac{1-\alpha ^2}{4989600 \pi ^2 \mu ^8 r^{16}}\sum^{6}_{j=0}\sum^{5}_{k=0}\chi_{jk}(\alpha, \mu)r^{2j}\xi^{k}
\label{phi4}
\end{equation}
were the coefficients $\chi(\alpha, \mu)$ that are non zero are given by
\begin{equation}
\chi_{00}(\alpha, \mu)=7810\left[ \alpha ^{12}+2\left( \alpha^{2}+\alpha ^{10}\right)-15\left( \alpha^{4}+\alpha ^8\right)+20 \alpha ^6\right]
\label{phi4coef00}
\end{equation}
\begin{equation}
\chi_{10}(\alpha, \mu)=4862649\alpha^{2}\{ -\alpha ^{12}+5\left[ \alpha ^{10}-2\left( \alpha^{6}+\alpha ^8\right)- \alpha ^4\right]+1\}
\label{phi4coef10}
\end{equation}
\begin{equation}
\chi_{20}(\alpha, \mu)=7461732 \alpha^{4}\left[ -\alpha ^{8}+4\left( \alpha^{2}+\alpha ^{6}\right)-6 \alpha ^4+1\right]
\label{phi4coef20}
\end{equation}
\begin{equation}
\chi_{30}(\alpha, \mu)=\alpha^{4}\left(-393139 \alpha ^{8}+5283387 \alpha ^{6}-4170957 \alpha ^4+1546549 \alpha ^2+2160\right)
\label{phi4coef30}
\end{equation}
\begin{equation}
\chi_{40}(\alpha, \mu)=\alpha ^{14} \mu ^2\left[106260\alpha^{4} \left(\alpha ^{2}+1\right)+23100 \alpha ^{2}+1980\right]-70272\alpha ^{10}\left(\alpha^{2}-1\right)
\label{phi4coef40}
\end{equation}
\begin{equation}
\chi_{50}(\alpha, \mu)=\alpha ^{14} \mu ^4\left[24750\left( \alpha ^{4}+\alpha^{2}\right)+3960\right] +492024 \alpha ^{10}\left(\alpha ^{2}-1\right)
\label{phi4coef50}
\end{equation}
\begin{equation}
\chi_{60}(\alpha, \mu)=20790 \alpha ^{14} \mu ^6\left(\alpha ^{2}+1\right) ,
\label{phi4coef60}
\end{equation}
\begin{equation}
\chi_{31}(\alpha, \mu)=\alpha^{4}\left(-7484400 \alpha ^{6}-2245320 \alpha ^4-237600 \alpha ^2-11880 \right)
\end{equation}
\begin{equation}
\chi_{41}(\alpha, \mu)=\alpha ^{14} \mu ^2\left(-831600 \alpha ^{4}-221760 \alpha ^{2}-15840\right) ,
\end{equation}
\begin{equation}
\chi_{51}(\alpha, \mu)=-41580 \alpha ^{14} \mu ^4\left(1+5 \alpha ^{2}\right)
\end{equation}
\begin{equation}
\chi_{61}(\alpha, \mu)=-207900 \alpha ^{14} \mu ^6,
\end{equation}
\begin{equation}
\chi_{32}(\alpha, \mu)=3960 \alpha ^4 \left(\alpha ^2-1\right) \left[616 \left(\alpha ^2-1\right) \alpha ^{14} \mu ^2-315 \left(18 \alpha ^6+9 \alpha ^4+\alpha ^2\right)-12\right]
\end{equation}
\begin{equation}
\chi_{42}(\alpha, \mu)=-27720 (\alpha -1) \alpha ^{14} (\alpha +1) \left(46 \alpha ^4+40 \alpha ^2+3\right) \mu ^2,
\end{equation}
\begin{equation}
\chi_{52}(\alpha, \mu)=-207900 \alpha^14 (-1 - 2 \alpha^2 + 3 \alpha^4) \mu^4
\end{equation}
\begin{equation}
\chi_{62}(\alpha, \mu)=623700 \alpha ^{14} \left(1-\alpha ^2\right) \mu ^6,
\end{equation}
\begin{equation}
\chi_{33}(\alpha, \mu)=-166320 \alpha ^4 \left(135 \alpha ^8-245 \alpha ^6+86 \alpha ^4+23 \alpha ^2-1\right)
\end{equation}
\begin{equation}
\chi_{43}(\alpha, \mu)=277200 \alpha ^{14} \left(1-\alpha ^2\right)^2 \left(8 \alpha ^2+1\right) \mu ^2,
\end{equation}
\begin{equation}
\chi_{53}(\alpha, \mu)= 415800 \alpha ^{14} \left(1-\alpha ^2\right)^2 \mu ^4
\end{equation}
\begin{equation}
\chi_{34}(\alpha, \mu)=415800 \alpha ^4 \left(1-\alpha ^2\right)^3 \left(15 \alpha ^2+1\right),
\end{equation}
\begin{equation}
\chi_{44}(\alpha, \mu)=415800 \alpha ^{14} \left(1-\alpha ^2\right)^3 \mu ^2
\end{equation}
and finally
\begin{equation}
\chi_{35}(\alpha, \mu)=498960 \alpha ^4 \left(1-\alpha ^2\right)^4
\end{equation}

\begin{figure} 
\centering
\includegraphics[width=10cm]{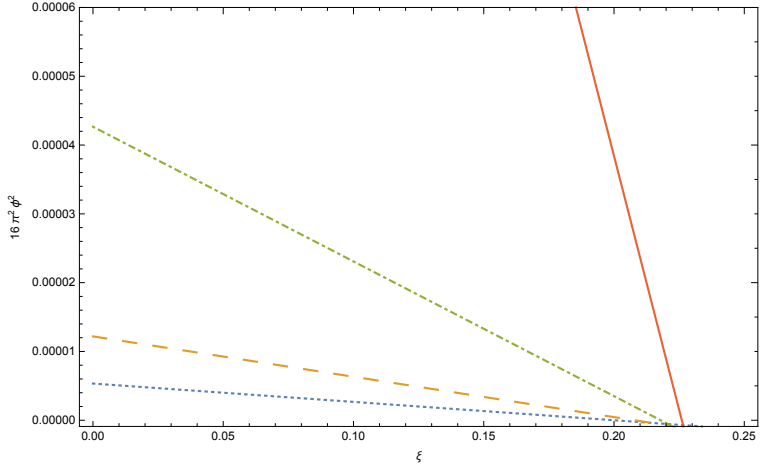}
\caption{ Dependance on the coupling constant $\xi$ of the field fluctuation $16\pi^{2}\left <\phi^{2}\right >^{\textit{(k)}}$ for $k=1$ (dotted line),\ $k=2$ (dashed line),\ $k=3$ (dotdashed line), \ and $k=4$ (solid line),  for massive scalar field in the pointlike global monopole spacetime. The values of the parameters used in the calculations are $\mu=2$, $r=\frac{1}{3}$, and $1 - \alpha^2 =10^{-5}$}
\label{f1}
\end{figure}

In the rest of the paper, we will analyze to what extent the inclusion of higher order terms in (\ref{main}) leads to appreciable corrections to the leading value $\left<\phi^{2}\right>^{(1)}$ of the vacuum polarization in the pointlike global monopole spacetime.

In Figure \ref{f1} we show the dependence of the four main approximations for the vacuum polarization on the coupling parameter $\xi$, for a scalar field with mass $\mu=2$ coupled to the point like global monopole at some fixed distance from the monopole's core.

The first thing to note is the almost linear dependance of the field fluctuation on the coupling parameter, in all the approximations. This fact is due to the small value of $\alpha^{2}$, which causes the terms proportional to $\xi^{k}$ with $k=2,3,4,5$ appearing in the expression for $\left <\phi^{2}\right >^{\textit{(k)}}$ to be very small, so that the mean effect is linear in $\xi$.

As we can easily seen from the figure, the four approximations leads to results that differ considerably from each other, with more pronounced differences for values of the coupling constant close to the minimal value $\xi=0$. This situation is in contrast with that encountered in other systems, as for example the Reissner-Nordstrom black hole spacetime \cite{matyjasek4}, for which the next to next to leading approximations gives results closely to the next to leading one, which indicates that $\left <\phi^{2}\right >^{\textit{(2)}}$ is a good approximation for the field fluctuation .

The surprisingly different result obtained in our case is an indication that, in the case of a pointlike global monopole spacetime, we need to take into account the next to next to next to leading term $\left <\phi^{2}\right >^{\textit{(4)}}$ to have a closer approximation to the real value of the field's fluctuation.

An interesting result that we can see from Figure \ref{f1} is that for each of the approximations, there exist some critical value of the coupling constant at which the field fluctuation vanishes, becoming negative for values greater than this critical value. However, the vacuum polarization is always positive for the minimal and the conformally coupled case.

In the following we shall confine ourselves to the physically important cases of minimally and conformally coupled scalar fields.

\subsection{Minimal coupling}

For the case of a massive scalar field minimally coupled to the gravitational field of the global monopole, the leading approximation for the field fluctuation is given by
\begin{equation}
\left <\phi^{2}\right >_{\textit{min}}^{\textit{(1)}}={\frac {1-{\alpha}^{4}}{240\,{\pi}^{2}{r}^{4}{\mu}^{2}}}
\end{equation}

The next to leading approximation results
\begin{equation}
\left <\phi^{2}\right >_{\textit{min}}^{\textit{(2)}}={\frac {21 \left(1-\alpha ^4\right) \mu ^2 r^2-25 \alpha ^6+21 \alpha ^2+4 }{5040\,{\pi}^{2}{\mu}^{4}{r}^{6}}}
\end{equation}

The next to next to leading term is given by (\ref{phi3}) with $j=1,2,3$ and the coefficients in (\ref{phi3coef10})-(\ref{phi3coef30}),
whereas the next to next to next to leading approximation is given by \ref{phi4} with $j=1$ to $j=6$ with coefficients given by formulas \ref{phi4coef00} to \ref{phi4coef60} respectively.
\begin{figure} 
\centering
\includegraphics[width=10cm]{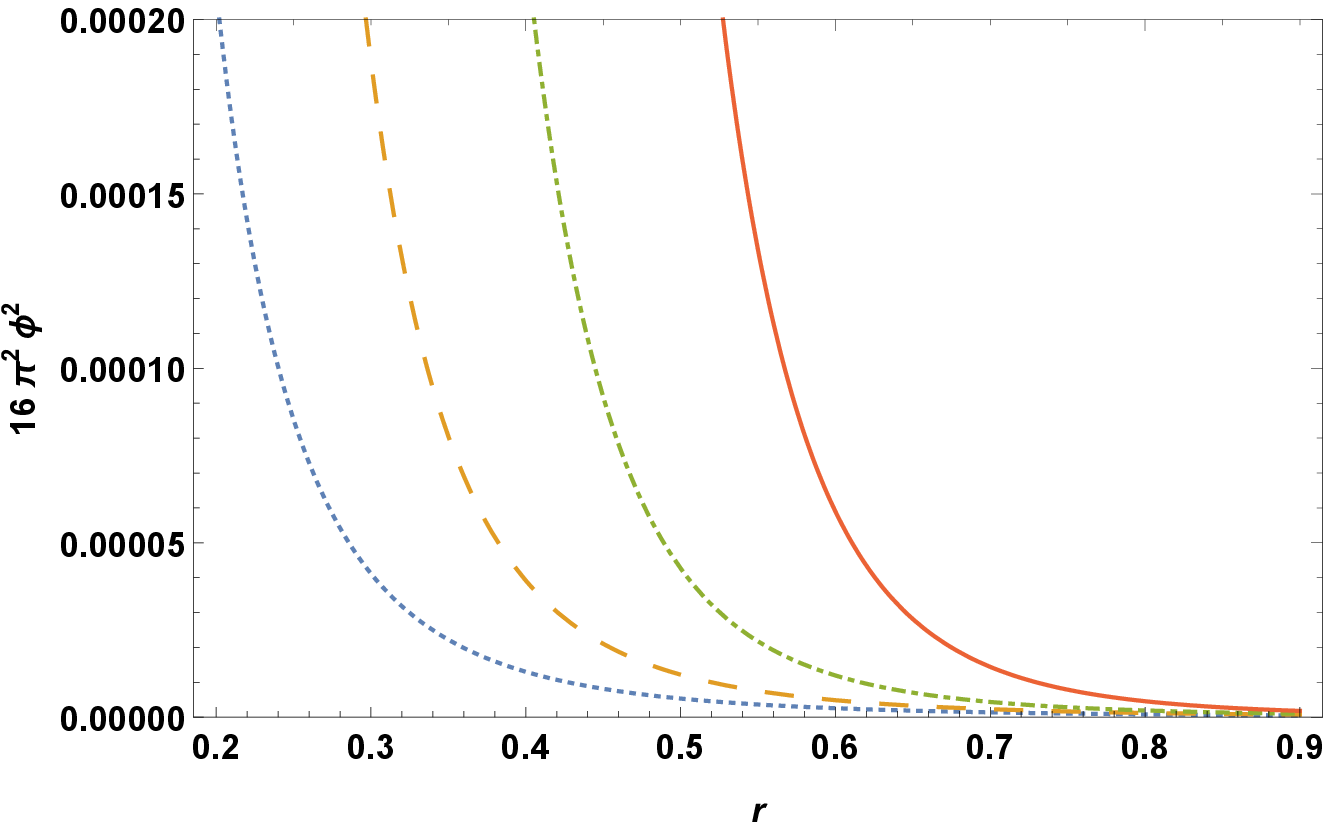}
\caption{ Dependance on the distance from the monopole of the field fluctuation $16\pi^{2}\left <\phi^{2}\right >_{\textit{min}}^{\textit{(k)}}$ for $k=1$ (dotted line),\ $k=2$ (dashed line),\ $k=3$ (dotdashed line), \ and $k=4$ (solid line),  for massive scalar field minimlly coupled to gravity in the pointlike global monopole spacetime. The values of the parameters used in the calculations are $\mu=2$, and $1 - \alpha^2 =10^{-5}$  }
\label{f2}
\end{figure}

In Figure \ref{f2} we show the dependance of the vacuum polarization with the distance $r$ to monopole's core, for the four main approximation studied in this paper. As we can see, the field fluctuation decreases from the monopole's core, with $\left <\phi^{2}\right >_{\textit{min}}^{\textit{(k)}}\rightarrow 0$ as $r\rightarrow \infty$.

The most important thing to be noted is the different results that produced all the considered approximation, because each term corresponding to a higher approximation considerably modifies the result of the previous one. This is similar to the result encountered for the dependance of the vacuum polarization with the coupling parameter $\xi$. This again is in contrast with previous calcultions in spherically symmetric spacetimes, specifically in the line element describing a Reissner-Nordstrom black hole, for which Matyjasek et. al find that the next to next to leading approximation differ only slightly from the next to leading one, an indication that the the changes in the field fluctuation produced by $[a_{4}]$ are relatively small \cite{matyjasek4}.

For the case of global monopole spacetime, it seems that the situation is completely different. Not only the next to next to leading term differ considerably from the next to leading one, also the same happens with the more higher next to next to next to leading term with respect to the previous one. All the above indicates that for the minimally coupled massive scalar field in the background of a pointlike global monopole we need to consider at most the first four terms in the expansion \ref{main},to provide a good approximation to the real value of the field fluctuation.

\subsection{Conformal coupling}

For conformal coupling we have $\xi=\frac{1}{6}$, which gives as the leading approximation to the vacuum polarization the result:
\begin{equation}
\left <\phi^{2}\right >_{\textit{conf}}^{\textit{(1)}}=\frac{1}{6}\left <\phi^{2}\right >^{\textit{(min)}}_{\textit{leading}}={\frac {1-{\alpha}^{4}}{1440\,{\pi}^{2}{r}^{4}{\mu}^{2}}}
\label{phi1conf}
\end{equation}
\begin{figure} 
\centering
\includegraphics[width=10cm]{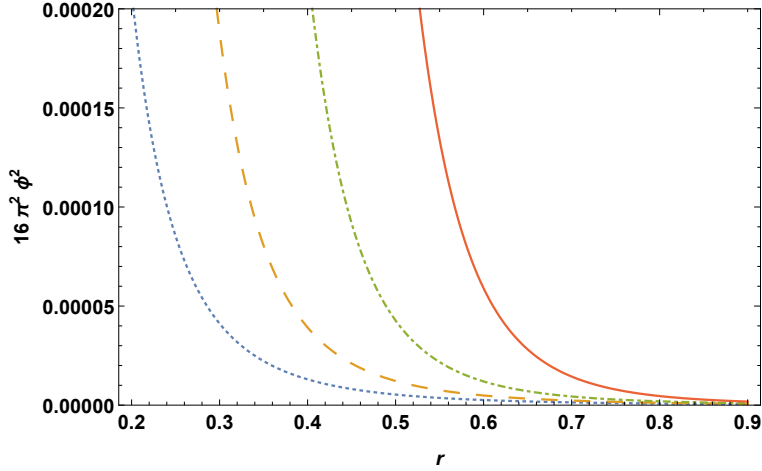}
\caption{ Dependance on the distance from the monopole of the field fluctuation $\Lambda=16\pi^{2}\left <\phi^{2}\right >_{\textit{conf}}^{\textit{(k)}}$ for $k=1$ (dotted line),\ $k=2$ (dashed line),\ $k=3$ (dotdashed line), \ and $k=4$ (solid line),  for massive scalar field conformally coupled to gravity in the pointlike global monopole spacetime. The values of the parameters used in the calculations are $\mu=2$, and $1 - \alpha^2 =10^{-5}$   }
\label{f3}
\end{figure}

For the next to leading approximation we obtain
\begin{equation}
\left <\phi^{2}\right >_{\textit{conf}}^{\textit{(2)}}={\frac { \left(1- {\alpha}^{2} \right)}{90720\,{\pi}^{2}{\mu}^{4}{r}^{6}}}\left[( {\alpha}^{2}+1)(63{\mu}^
{2}{r}^{2}+100\,{\alpha}^{2})+16
 \right]
 \label{phi2conf}
\end{equation}

The next to next to leading approximation is given by
\begin{equation}
\left <\phi^{2}\right >_{\textit{conf}}^{\textit{(3)}}=\frac { \left(1- {\alpha}^{2} \right)}{18144\,{r}^{10}{\pi}^{2}{
\mu}^{6}}\sum_{k=0}^{3}q_{k}(\alpha)\, r^{2k}\,\mu^{2(k-1)}
\end{equation}
where $q_{0}(\alpha)=2464\,{\alpha}^{4}(\,{\alpha}^{4}-2\,{\alpha}^{2}+1)$\,$q_{1}(\alpha)=2191\,{\alpha}^{6}-273\,{\alpha}^{4}+217\,{\alpha}^{2}+32$ \, $q_{2}(\alpha)=200\,{\alpha}^{4}+200\,{\alpha}^{2}+32$\, and $q_{3}(\alpha)=126\,({\alpha}^{2}+1)$.

The next to next to next to leading term for the field fluctuation for conformal coupling is
\begin{equation}
\left <\phi^{2}\right >_{\textit{conf}}^{\textit{(4)}}=\frac{1-\alpha ^2}{4989600 \pi ^2 \mu ^8 r^{16}}\sum_{k=0}^{6}\Lambda_{k}(\alpha,\mu)\, r^{2k}
\end{equation}
where
\begin{equation}
\Lambda_{0}(\alpha,\mu)=7810 \left(\alpha ^2-1\right)^6,
\end{equation}
\begin{equation}
\Lambda_{1}(\alpha,\mu)=-4862649 \alpha ^2 \left(\alpha ^2-1\right)^5,
\end{equation}
\begin{equation}
\Lambda_{2}(\alpha,\mu)=7461732 \alpha ^4 \left(\alpha ^2-1\right)^4,
\end{equation}
\begin{IEEEeqnarray}{rCl}
\Lambda_{3}(\alpha,\mu)&=&\frac{1}{3} \alpha ^4 \left[-3376997 \alpha ^8+13652581 \alpha ^6-13044171 \alpha ^4+4580027 \alpha ^2 \right.\nonumber\\
&&\left. +203280 \left(\alpha ^2-1\right)^2 \alpha ^{14} \mu ^2+2960\right],
\end{IEEEeqnarray}
\begin{equation}
\Lambda_{4}(\alpha,\mu)=\frac{55}{2} \alpha ^{14} \left[7 \left(313 \alpha ^4-39 \alpha ^2+31\right) \alpha ^2+25\right] \mu ^2-70272 \alpha ^{10} \left(\alpha ^2-1\right),
\end{equation}
\begin{equation}
\Lambda_{5}(\alpha,\mu)=492024 \left(\alpha ^2-1\right) \alpha ^{10}+220 \left(25 \left(\alpha ^4+\alpha ^2\right)+4\right) \alpha ^{14} \mu ^4
\end{equation}
and
\begin{equation}
\Lambda_{6}(\alpha,\mu)=3465 \alpha ^{14} \left(\alpha ^2+1\right) \mu ^6
\end{equation}

From the above expressions, we see that for the conformally coupled case, the field fluctuation is also a decreasing function of the distance $r$ from monopole's core, vanishing for large values of $r$. A strong similarity with the minimally coupled case is appreciated from Figure \ref{f3}, when we plot the quantity $\Lambda=16\pi^{2}\left <\phi^{2}\right >_{\textit{conf}}^{\textit{(k)}}$ as a function of $r$ using the values $\mu=2$, and $1 - \alpha^2 =10^{-5}$ for the parameters involved in the calculations.

As we can observe, a conformally coupled massive scalar field have field fluctuation strongly dependent of the approximation used. Again the approximation $\left <\phi^{2}\right >_{\textit{conf}}^{\textit{(4)}}$ differ considerably from the next to next to leading term $\left <\phi^{2}\right >_{\textit{conf}}^{\textit{(3)}}$, and this two approximations also differs from the leading and the next to leading ones. This is a strong indication of the necessity to take into account the coincidence limits of the Hadamard-DeWitt coefficients up to $\left[a_{5}\right]$, to get a good approximation for the vacuum polarization.

\section{Concluding remarks}

In this paper we set ourselves the objective of constructing the approximate field fluctuations, for a quantized massive scalar field in the space-time of a pointlike global monopole. Using the Schwinger-DeWitt expansion for the Green's function associated with the Klein-Gordon dynamical operator, we find analytical expressions for the vacuum polarization, as an expansion in powers of inverse squared mass of the field.

The different approximations constructed go from the leading one, proportional to $\mu^{-2}$, being $\mu$ the mass of the scalar field, up to terms involving the powers $\mu^{-8}$, which form which we called the next to next to next to leading approximation for $\left<\phi^{2}\right>$. The leading term is proportional to the coincidence limit of the Hadamard-DeWitt coefficient $\left[a_{2}\right]$, whereas the higher order term considered in this work is proportional to $\left[a_{5}\right]$.

Although the Hadamard-DeWitt coefficients are extremely complicated local expressions constructed from the Riemmann tensor, their covariant derivatives, and contractions, in this work we show that, in the case of spacetime associated with a global monopole, these expressions are considerably simplified, giving place to relatively simple results.

The results obtained for the field fluctuations of the quantized massive scalar field in the global monopole background shows that taking into account higher order terms substantially improve the approximation. We concluded that for this spacetimes, we need to use the next to next to next to leading term to obtain a good description of the vacuum polarization.

This situation is in contrast with that obtained for other spacetimes with spherical symmetry, as the one describing a Reissner-Nordstrom black hole, for which previous studies showed that the next to leading term, proportional to $\left[a_{3}\right]$, provides a reasonable good approximation.

Our calculations constitutes a first step in the study of vacuum polarization of massive fields in pointlike global monopole's background, using the Schwinger-DeWitt proper time formalism. Future work will be devoted to the determination of the renormalized stress-energy tensor for the quantized field in this spacetime. Finding the stress energy tensor involves the calculation of the functional derivatives of the renormalized quantum effective action, whose leading term is proportional to the coincidence limit of the Hadamard-DeWitt coefficient $\left[a_{3}\right]$. For this reason, it is a more difficult problem than the related calculation of the field fluctuation.

However, for conformally coupled scalar field, there exist a relation between the field fluctuation of the scalar field and the trace anomaly of the stress energy tensor. As showed by Anderson in reference \cite{anderson}, we have for the trace $\left<T_{\nu}^{\nu}\right>$ the expression
\begin{equation}
\left<T_{\nu}^{\nu}\right>=\frac{a_{2}}{16 \pi^{2}}-\mu^{2}\left<\phi^{2}\right>
\label{tanomaly}
\end{equation}
As the leading approximation for the field fluctuation exactly cancel the anomaly in \ref{tanomaly}, proportional to $\left[a_{2}\right]$, then the leading approximation to the trace of the quantum stress energy tensor comes from the next to leading term in the field fluctuation, that involves $\left[a_{3}\right]$.

Substituting \ref{phi1conf} and \ref{phi2conf} in \ref{tanomaly} we obtain for the trace of the quantum stress energy tensor for a conformally coupled massive scalar field in the pointlike global monopole spacetime the result
\begin{equation}
\left<T_{\nu}^{\nu}\right>=\frac{25 \alpha ^6-21 \alpha ^2-4}{22680 \pi ^2 \mu ^2 r^6}
\label{tanomalymonopole}
\end{equation}

This derivation of the above result involve much less effort in comparison with the calculation of the components of the stress energy tensor by functional derivation of the effective action with respect to the metric tensor, and the subsequent summation of diagonal entries of this quantity, and \ref{tanomalymonopole} can be useful to check the results of that type of calculation.

Another interesting problem arise when considering the backreaction effects of the quantized massive fields upon the spacetime geometry around the monopole. In future reports we will address also this specific problems.

\section*{Acknowledgments}
This work has been supported by TWAS-CONACYT $2017$ fellowship, that allow to the author to do a sabatical leave at Departamento de F\'isica Te\'orica, Divisi\'on de Ciencias e Ingenier\'ias, Universidad de Guanajuato, Campus Le\'on. The author also express his gratitude to Professor Oscar Loaiza Brito, for the support during the research stay at his group, where this work was completed.


\end{document}